# Student thinking about the divergence and curl in mathematics and physics contexts

Charles Baily,[1] Laurens Bollen,[2] Andrew Pattie,[1] Paul van Kampen[3] and Mieke De Cock[2]

[1]*School of Physics and Astronomy, University of St Andrews,*
*North Haugh, St Andrews, Fife KY16 9SS Scotland, UK*
[2]*Department of Physics and Astronomy & LESEC, KU Leuven*
*Celestijnenlaan 200c, 3001 Leuven, Belgium*
[3]*Centre for the Advancement of Science and Mathematics Teaching and Learning*
*& School of Physical Sciences, Dublin City University, Glasnevin, Dublin 9, Ireland*

**Abstract.** Undergraduate physics students are known to have difficulties with understanding mathematical tools, and with applying their knowledge of mathematics to physical contexts. Using survey statements based on student interviews and written responses to open-ended questions, we investigated the prevalence of correct and incorrect conceptions regarding the divergence and curl of vector fields, among both mathematics and physics students. We compare and contrast pre-instruction responses from intermediate-level E&M students at KU Leuven and the University of St Andrews, with post-instruction responses from St Andrews students enrolled in a vector calculus course. The differences between these student populations were primarily in areas having to do with physics-related concepts and graphical representations of vector fields. Our comparison of pre- and post-instruction responses from E&M students shows that their understanding of the divergence and curl improved significantly in most areas, though not as much as would be desired.

## I. INTRODUCTION

Vector calculus is an important mathematical tool employed by physicists in the study of a variety of phenomena. This is particularly true in electromagnetism (E&M), where misunderstandings of what the divergence and curl of a vector field represent may lead to any number of student difficulties with the topic. It is not uncommon for undergraduate physics students to first learn about vector calculus in courses taught by mathematicians; depending on the institution, they may also (or instead) encounter it in a mathematical methods course, before ever learning to apply it in a physics context. Thus, upon entering an intermediate-level E&M course, students often have become adept at calculating the divergence or curl when given an explicit expression for a vector field, yet still exhibit a variety of misconceptions about their meaning, and what they physically represent. [1-4]

Dray and Manogue have previously drawn attention to the disparity in how mathematicians typically teach vector calculus and how it is used by physicists, and have discussed anecdotally some of the implications this "vector calculus gap" may have for student learning. [5] Others have shown that students' difficulties with interpreting graphical representations of solenoidal and irrotational fields can extend into their graduate careers, [6] which suggests that a conceptual understanding of vector calculus can remain elusive even for high-performing students.

As a step towards designing effective instructional interventions, it is important to first explore and document the variety of student thinking associated with the divergence and curl of a vector field. To this end, we have administered a research-based survey at two different institutions in two different countries, in both mathematics and physics classrooms, to gauge the relative prevalence of correct and incorrect ideas students have about these mathematical tools, and how those ideas changed in response to instruction. We have found notable differences between responses from students in intermediate-level E&M courses and students in a mathematics course on vector calculus. We also see that many physics students continue to hold incorrect ideas about the divergence and curl despite explicit instruction.

## II. METHODOLOGY

As part of a larger study of student difficulties with mathematics in electromagnetism [3], we sought to identify students' *concept image* [7] of the gradient, divergence and curl through a written free-response test given to E&M students at KU Leuven (N = 30) and Dublin City University (N = 50). One of the open-ended questions asked them to write down "everything you think of when you see $\nabla A$, $\nabla \cdot \mathbf{A}$ and $\nabla \times \mathbf{A}$." We found that only ~20% of these students provided a conceptual explanation, and that their descriptions were often quite incomplete and/or incorrect. This was followed by eight semi-structured interviews with individual students at KU Leuven, wherein student thinking about the divergence and curl was explored in greater detail.

Results from the written responses and interviews were used to assemble two lists of statements representative of common student thinking regarding the divergence and curl. These lists form the basis of the first two questions on a pre- and post-instruction survey designed to assess students' conceptual understanding of the divergence and curl, their ability to interpret graphical representations of vector fields, and to draw connections between the two. The prompts for these two questions ask students to "[I]ndicate which statements, formulas or properties are always correct when describing the curl [divergence] of a vector field." [See Table 1.]

**TABLE 1.** List of statements used in the pre- and post-instruction surveys regarding the curl (top panel) and divergence (bottom panel). Correct statements are highlighted in bold.

| | |
|---|---|
| C1 | $\nabla A$ |
| C2 | $\nabla \cdot \mathbf{A}$ |
| **C3** | $\nabla \times \mathbf{A}$ |
| C4 | The curl indicates where field lines start or end. |
| C5 | The curl is a measure for how much field lines bend. |
| C6 | The curl points in the direction of steepest increase. |
| **C7** | **The curl has a direction.** |
| C8 | The curl is nonzero if and only if the direction of the field changes. |
| **C9** | **The curl is a measure of the infinitesimal rotation of the field.** |
| C10 | The curl is a characteristic of the field, and is the same everywhere in the field. |
| **C11** | **The curl is represented by a vector.** |
| C12 | The curl is represented by a scalar. |
| D1 | $\nabla A$ |
| **D2** | $\nabla \cdot \mathbf{A}$ |
| D3 | $\nabla \times \mathbf{A}$ |
| **D4** | **The divergence measures the source or sink of a vector field.** |
| D5 | The divergence is a measure for how much field lines spread apart. |
| D6 | The divergence points in the direction of steepest increase. |
| D7 | The divergence has a direction. |
| D8 | The divergence is a measure for how much field lines bend. |
| **D9** | **The divergence indicates where field lines start or end.** |
| **D10** | **The divergence can be different for every spot in the field.** |
| D11 | The divergence is represented by a vector. |
| **D12** | **The divergence is represented by a scalar.** |

As suggested by Dray and Manogue, we suspected there might be differences in how physics students and mathematics students think about the divergence and curl of a vector field, but had nothing other than anecdotal evidence for this. We also wished to explore the impact of learning about and using vector calculus in the context of undergraduate E&M. We therefore gave the pre-survey at the start of the semester to intermediate E&M students at KU Leuven and the University of St Andrews, and also at mid-semester to students enrolled in a vector calculus course offered by the mathematics department in St Andrews. The post-survey was given at mid-semester to the same E&M students at KU Leuven and St Andrews, after they had covered material up through the full Maxwell equations in differential form.

## III. RESULTS AND DISCUSSION

The E&M courses at KU Leuven and St Andrews are taught at the level of Griffiths, [8] and cover both statics and electrodynamics in a single semester. At the time of their enrollment, all of the KU Leuven students had taken at least one math course that included vector calculus. Approximately 2/3 of the St Andrews E&M students first learned about vector calculus in a math methods course offered by the physics department in the previous semester (see Ref. [2]). About 1/3 had taken a semester of vector calculus in the previous year; some of these students took both courses. The pre-instruction responses for the E&M students at St Andrews (N = 71) and KU Leuven (N = 33) are statistically similar ($p = 0.43$ by a chi-square test), and we have combined them for the purposes of this discussion

The vector calculus course at St Andrews is taught in a traditional ("chalk and talk") large-lecture format, with a student population comprised of approximately 2/3 math majors, and 1/3 physics majors or physics & math double-majors. There was little time devoted during lecture to addressing common student difficulties, or engaging in sense making. The divergence and curl were taught without much use of visual representations of vector fields, and with little discussion of how vector calculus is applied in physical contexts.

### A. Comparisons between math and physics students

We first compare and contrast aggregate responses from the E&M students at pre-instruction (N = 104, in total) with those from the post-instruction vector calculus students at St Andrews (N = 124), as shown in Fig. 1. The distributions of responses from each of the two groups are statistically different, for both the curl and divergence ($p < 0.001$ by a chi-square test).

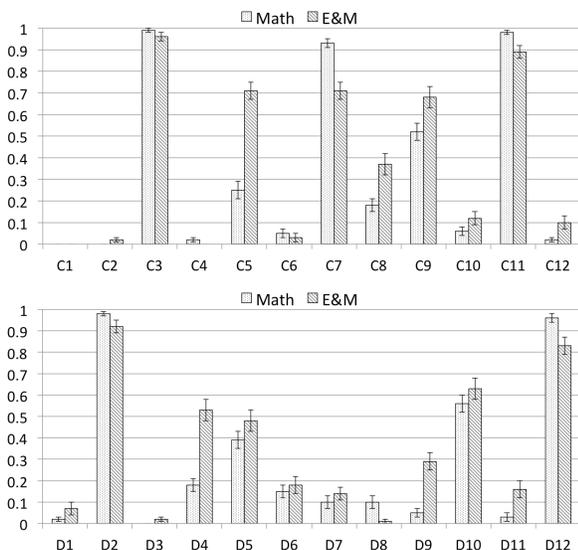

**FIG 1.** Fraction of post-instruction vector calculus students ("Math", $N = 124$) and pre-instruction E&M students ($N = 104$) selecting as correct statements C1–C12 (top panel) and D1–D12 (bottom panel). Error bars represent the standard error on the proportion. Correct responses are: C3, C7, C9 & C11; D2, D4, D9, D10 & D12.

Although it was not uncommon for students to write down incorrect mathematical expressions for the divergence and curl during interviews, or to incorrectly name the expressions $\nabla \cdot \mathbf{A}$ and $\nabla \times \mathbf{A}$ in the free-response survey, almost every student was able to select a correct expression for the curl and divergence (C3 & D2). Also contrary to our earlier observations, students performed relatively well at identifying the curl and divergence as vector or scalar quantities (C11 & D12); notably, the results for math students on these statements were somewhat better than for the E&M students ($p < 0.01$ for both C11 & D12 by a two-tailed t-test).

We had also anticipated from previous results that more students would mistakenly associate properties of the gradient with the divergence or curl [C/D6: "The curl (divergence) points in the direction of steepest increase."]. Hardly any students attributed to the curl a property that correctly belongs to the divergence [C4: "The curl indicates where field lines start or end."]; nor did they attribute to the divergence a property that students sometimes (incorrectly) associate with the curl [D8: "The divergence is a measure for how much field lines bend."]. All together, these results suggest that much of the confusion and incompleteness displayed in students' free responses to a written questionnaire essentially disappeared when the task involved selecting correct statements from a list that had been generated for them.

Other statements reveal some apparent inconsistencies in student thinking. Few students were distracted by C10 ["The curl is a characteristic of the field, and is the same everywhere in the field."], yet only ~60% of all students indicated that D10 is correct ["The divergence can be different for every spot in the field."]. Even though most students identified the curl as a vector quantity, we note that 23% of the E&M students who selected C11 did not also select C7 ["The curl has a direction."], whereas the consistency on these two statements was much greater for the math students.

The remaining statements reveal other disparities between the two student groups, particularly those having to do with visual representations of vector fields, or concepts that are more physics related. D5 and C5 reflect common incorrect associations of a non-zero divergence (curl) with regions in a diagram where field lines spread apart (bend). The results are statistically similar for both populations regarding the divergence ($p = 0.16$); however, 71% of E&M students selected statement C5, but only 25% of math students. E&M students performed better than math students on D9 ["The divergence indicates where field lines start or end"; 29% vs. 5%], and also on D4 ["The divergence measures the source or sink of a vector field; 53% vs. 18%.]. The curl as a measure of the infinitesimal rotation of a vector field (C9) is a fairly mathematical definition, yet E&M students were more likely than math students to select this as a correct statement (68% vs. 52%; $p = 0.01$).

### B. Pre- and post-instruction E&M comparisons

The distribution of post-instruction responses for the E&M students at St Andrews ($N = 71$) and at KU Leuven ($N = 15$) are statistically similar ($p = 0.27$ by a chi-square test), and we have again combined them for the purpose of making comparisons between their mid-semester responses and those at pre-instruction (see Fig. 2). It is immediately apparent that the E&M students showed significant gains on almost every statement where there was room for substantial improvement.

Students continued to perform well at identifying mathematical expressions for the divergence and curl, and on statements C/D11-12 (divergence and curl as scalar and vector quantities). The percentage of students selecting incorrect statements either remained low (C4, C6, C10 & D6-8), or reduced substantially to around 10% or less (C5, C8 & D5). Significant improvements can also be seen in the percentage of students who selected correct statements C7 (from 71% to 94%), D4 (from 53% to 85%), D9 (from 29% to 71%) and D10 (from 63% to 81%); $p \leq 0.005$ for all four statements. However, the percentage of E&M students correctly choosing statement C9 ["The curl is a measure of the infinitesimal rotation of the field."] actually decreased at post-instruction (from 68% to 59%), though the shift is not statistically significant ($p = 0.21$ by a two-tailed t-test).

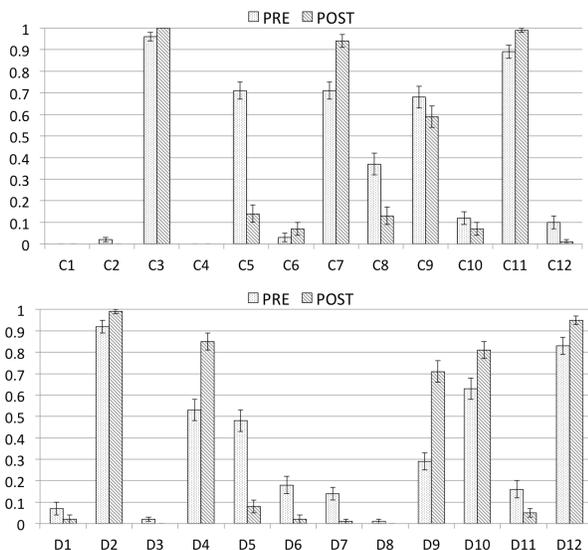

**FIG 2.** Fraction of E&M students selecting statements C1–C12 (top panel) and D1–D12 (bottom panel) at pre-instruction (N = 104) and post-instruction (N = 86). Error bars represent the standard error on the proportion. Correct responses are: C3, C7, C9 & C11; D2, D4, D9, D10 & D12.

## IV. CONCLUSIONS

Responses to our research-based survey have been used to gauge the relative prevalence of specific ideas students have about the divergence and curl of a vector field, whether correct or incorrect. These results demonstrate significant differences in answers given by E&M students at pre-instruction and vector calculus students at post-instruction, primarily in areas having to do with physics-related concepts and graphical representations of vector fields. We also see that explicit instruction has been moderately effective in shifting students towards a more normative concept image of the divergence and curl, but that there is still room for improvement in helping students to develop their understanding.

The observed decrease in the percentage of correct responses from E&M students on statement C9 suggests that the ways in which E&M is being taught at St Andrews and KU Leuven do not reinforce this concept. Although students may have learned that current densities and time-varying electric fields are sources of magnetic fields (and time-varying magnetic fields for non-conservative E-fields), many of them apparently did not draw connections between the mathematical content in the definition of the curl, and the physical content of Faraday's law and the Maxwell-Ampere law.

The majority of the differences between pre-instruction E&M students and post-instruction math students were related to visual representations of vector fields, and the relationship between a field's divergence and/or curl and the location of field sources. To further explore possible explanations for these differences, we separated the responses from the vector calculus students into two groups: those from pure math majors ($N = 79$), and those from students who were concurrently taking a second-year physics course ($N = 36$) (the E&M portion of this course uses Maxwell's equations in integral form). There were apparent differences between these two subgroups on the same statements as for the pre-E&M students and post-Math students in aggregate, but in this case the differences were not statistically significant ($p = 0.42$ for the divergence statements, and also for the curl statements). Further data collection will be necessary to more robustly establish any differences between these two student populations.

We also wish to investigate some of the possible origins of the more common incorrect ideas students have regarding the divergence and curl. For example, we suspect the (incorrect) notion that the divergence (curl) of a field is either zero everywhere or non-zero everywhere in space may stem from the typical diagrams used to illustrate their geometrical interpretations (e.g., see Figs. 1.18-19 in Ref. [8]). A contributing factor may be the usual types of vector fields (more specifically, the mathematical expressions) seen by students when first learning to calculate a divergence or curl. Presumably, mathematics students rarely (if ever) work with piecewise definitions of vector functions in different regions of space, as when considering the magnetic field due to a current-carrying wire.

Based on the results reported here, we will be conducting additional student interviews in order to more deeply explore student thinking, and to further validate the survey statements. The outcomes from these studies are being used to develop and evaluate research-based learning materials that target known student difficulties with mathematics in the context of electromagnetism.

## ACKNOWLEDGEMENTS

This work was supported in part by the University of St Andrews, KU Leuven and Dublin City University. The authors wish to thank the instructors and students whose cooperation and participation made these studies possible.